\documentclass[twocolumn,showpacs,preprintnumbers,prl]{revtex4}
\usepackage{amsfonts}
\usepackage{amsmath}
\usepackage{graphicx}
\usepackage{dcolumn}
\usepackage{bm}

\begin{document}

\title{Dynamics of quantum collapse in coupled quantum dots}
\author{H. Cruz}
\email{hcruz@ull.es}
\affiliation{\textit{Departamento de F\'{\i}sica, Universidad de La Laguna,
38204 La Laguna, Tenerife, Spain}}
\date{\today }

\begin{abstract}
In this letter, we have considered an electron in a coupled quantum dot
system interacting with a detector represented by a point contact. We
present a dynamical model for wave function collapse in the strong coupling
to the detector limit. In our model, the electron in the double quantum dot
makes a fast transition minimizing the emission of electromagnetic
radiation. In this way, a principle of leats emitted radiation can provide a
possible description of wave function collapse.
\end{abstract}

\maketitle


\bigskip The problem of understanding whether a measurement process can be
analyzed within the quantum mechanical formalism has long been a difficult
unresolved issue in the foundations of quantum mechanics. On one hand, the
quantum theory states that the vector corresponding to a physical system
undergoes a continuos evolution governed by Schr\"{o}dinger equation; on the
other hand, the theory prescribes a sudden jump motion to the state of a
physical system undergoing a measurement by an external device. Von
Neumann's projection rules \cite{von} are indeed to be added to the quantum
formalism in order to account for the transition from a pure to a mixed
state (the so-called wave function collapse), and this makes quantum
mechanics a non-self-contained theory. The renewed interest in the
measurement problem is justified by the development of mesoscopic systems
sensitive to the phase of the electronic wave function. Recent proposals
suggested using mesoscopic devices, such a Josephson junctions or coupled
quantum dots, as quantum bits (qubits), which are the basic elements of
quantum computers \cite{joseph}. Among various modern approaches to the
measurement problem in mesoscopic structures let us mention the idea of
replacing the collapse postulate by the gradual decoherence of the density
matrix due to the interaction with the detector \cite{zurek} and the
approach of a stochastic evolution of the wave function \cite{korotkov}.

Recently, Gurvitz \ \textit{et . al.} \cite{gurvitz} have considered a qubit
interacting with its environment and continuously monitored by a detector
represented by a point contact. In such a case, the decoherence rate $\Gamma
_{d}$ due to interaction with the detector is inversely proportional to the
measurement time $\Delta t$. For strong coupling to the detector, i.e., $%
\Gamma _{d}\rightarrow \infty $ and $\Delta t\rightarrow 0$, the measurement
is idealized to be instantaneous. Accordingly, the electron in the coupled
quantum dot system instantaneously makes the transition $|\psi
_{i}>\rightarrow |\psi _{f}>$ by measurement (the so-called quantum-jump 
\cite{milburn}). However, we notice that one remaining key question in these
theories is theoretical analysis of electromagnetic radiation emitted by a
charged particle during collapse. In a nonrelativistic collapse model,
spontaneous radiation of free electrons has been recently studied by Fu \cite%
{fu}. If the wave function collapse is instantaneous, the electromagnetic
radiation emitted by a charged particle diverges. This implies an infinite
value for the radiation reaction force. By contrast, an infinite value for
the emitted radiation can be eliminated by considering a dynamical model for
wave function collapse. This is the aim of the present work. In this letter,
we present a dynamical model for wave function collapse in the strong
coupling limit. In our model, the particle makes the transition $|\psi
_{i}>\rightarrow |\psi _{f}>$ as soon as possible but not instantaneously.
We assume that the electron in the coupled quantum dot makes a fast
transition minimizing the emission of electromagnetic radiation. In this
way, a principle of leats emitted radiation can provide a possible
description of wave function collapse:%
\begin{equation}
\delta \int_{t_{i}}^{t_{f}}I(t)dt=0\text{,}  \label{principle}
\end{equation}%
being $I(t)$ the intensity of radiation per unit time and $t_{i}$, $t_{f}$
the initial and final time, respectively. A leats emitted radiation
principle will give us the minimum possible collapse times without radiation
reaction divergences. Such a collapse time value will correspond to a
minimum possible measurement time value in a double quantum dot system.

Let us now consider electrostatic quantum bit measurements in double quantum
dots \cite{meas}. The qubit is a single electron and the detector is a point
contact placed near one of the dots \cite{butt}. If the electron occupies
the first dot, the transmission coefficient of the point contact decreases
due to electrostatic repulsion generated by the electron. Thus, the electron
position is monitored by the tunneling current. The qubit can be described
by the Hamiltonian \cite{gurvitz}

\begin{equation}
H=E_{l}a_{l}^{\dagger }a_{l}+E_{r}a_{r}^{\dagger }a_{r}+t(a_{l}^{\dagger
}a_{r}+a_{r}^{\dagger }a_{l})  \label{sch}
\end{equation}%
where $a_{l}^{\dagger }$($a_{l}$) and $a_{r}^{\dagger }$($a_{r}$)\ are the
creation (annihilation) operators in the left and rigth quantum wells and $t$
is the hopping amplitude between states $|l>$ and $|r>$ of the coupled
quantum dot. The coupled quantum dot potential has alternating odd and even
pure eigenstates ($|->$ and $|+>$). Superposition of these eigenstates can
be constructed so as to give states well localized in the left or right
well. Then, the ($|l>$ and $|r>$) states are given by

\begin{equation}
|l>=\frac{1}{\sqrt{2}}(|+>+|->)  \label{states}
\end{equation}%
and by

\begin{equation}
|r>=\frac{1}{\sqrt{2}}(-|+>+|->)\text{,}  \label{states2}
\end{equation}%
respectively. The system under observation is in a pure quantum state $|\psi
(t)>$ at the beginning of the measurement. Then, it will be in a pure
conditional state after the measurement, conditioned on the result. If the
initial state is $|\psi _{i}>=|+>$, the unnormalized final state given the
result $|\psi _{f}>=|r>$ at the end of the measurement becomes $|\psi
_{f}>=(|r><r|)|\psi _{i}>$, where \{$|l><l|$, $|r><r|$ \} represents a set
of operators that define the measurements and satisfies%
\begin{equation}
|l><l|+|r><r|=1\text{.}  \label{proy}
\end{equation}%
When this occurs, the system undergoes an instantaneous evolution $%
|+>\rightarrow |r>$, called a quantum jump.

Let us now consider a principle of leats emitted radiation, Eq. (\ref%
{principle}). From quantum electrodynamics we know that the intensity of
dipole radiation is \cite{landau} 
\begin{equation}
I(t)=\frac{1}{4\pi \epsilon }\frac{2}{3c^{3}}|\ddot{p}(t)|^{2}\text{,}
\label{dipole}
\end{equation}%
being $c$ the speed of light, $p$ the dipole moment and $\epsilon $ the
dielectric constant. This is directly analogous to the classical formula for
the intensity of dipole radiation from a system of moving particles. We note
the correspondence principle for the radiation intensity is valid not only
in the quasi-classical but in the general quantum case \cite{landau}. In the
one-dimensional case, the dipole moment $p(t)$ of the electron in the
coupled quantum dot is%
\begin{equation}
p(t)=e  x |\psi (x,t)|^{2}dx=e<x(t)>  \label{p}
\end{equation}%
where $e|\psi (x,t)|^{2}$ is the electron charge density. In the pure
dipolar case, Eq. (\ref{principle}) can be written as

\begin{equation}
\delta \int_{t_{i}}^{t_{f}}\left\vert \frac{\partial ^{2}}{\partial t^{2}}%
<x>\right\vert ^{2}dt=0\text{.}  \label{p2}
\end{equation}%
Now we solve Eq. (\ref{p2}) for a double quantum dot. We consider the
transition $|+>\rightarrow |r>$ by measurement (Fig. 1). In such a case, and
during the collapse, $<x>$ is given by $<x>=[1-A(t)]x_{l}+A(t)x_{r}$, where $%
<l|x|l>=x_{l}$, $<r|x|r>=x_{r}$ and $A(t)$ is a time-dependent coefficient ($%
0\leq A(t)\leq 1$). For simplicity, let us assume that the coordinate origin
is placed in the middle of the left quantum well, i.e., $x_{l}=0$. We call $%
A(t)$\ the true path and we take some trial path $A_{t}(t)$ that differs
from the true path by a small amount which we will call $\eta (t)$. We
calculate the emitted radiation $E_{rad}$ for the path $A(t)$ and $A(t)+\eta
(t)$, respectively. The difference must be zero in the first-order
approximation of small $\eta (t)$. We rearrange Eq. (\ref{p2}) by
integrating by parts and considering $\eta (t_{i})=\eta (t_{f})=0$ and $\dot{%
\eta}(t_{i})=\dot{\eta}(t_{f})=0$. Leaving out second and higher order
terms, we have for $\delta E_{rad}$ 
\begin{equation}
\delta E_{rad}=\int_{t_{i}}^{t_{f}}\eta (t)\ddddot{A}(t)dt=0\text{.}
\label{delta-e}
\end{equation}%
This means that the function $\ddddot{A}(t)$ is zero. Equation $\ddddot{A}%
(t)=0$ can be easily solved considering $A(t_{i})=1/2$, $A(t_{f}-t_{i})=1$
and $\dot{A}(t_{i})=\dot{A}(t_{f}-t_{i})=0$. Then, and during the
measurement process, we have for $|\psi (t)>$

\begin{equation}
|\psi (t)>=\left( \sqrt{\frac{t^{3}}{\tau _{c}^{3}}-\frac{3t^{2}}{2\tau
_{c}^{2}}+\frac{1}{2}}\right) |l>-e^{i\phi }\left( \sqrt{-\frac{t^{3}}{\tau
_{c}^{3}}+\frac{3t^{2}}{2\tau _{c}^{2}}+\frac{1}{2}}\right) |r>  \label{psi}
\end{equation}%
where $\phi $ is an arbitrary phase factor ($\phi =0$ at $t=0$) and $\tau
_{c}=t_{f}-t_{i}$ \ is the \textit{collapse time}. We define the operator $%
C(t)$ that verifies $|r>=C(t)|+>$. Taking into account Eq. (\ref{psi}), we
have 
\begin{equation}
C(t)=\left( 
\begin{array}{cc}
\sqrt{\frac{2t^{3}}{\tau _{c}^{3}}-\frac{3t^{2}}{\tau _{c}^{2}}+1} & 0 \\ 
0 & -e^{i\phi }\sqrt{-\frac{2t^{3}}{\tau _{c}^{3}}+\frac{3t^{2}}{\tau
_{c}^{2}}+1}%
\end{array}%
\right) \text{,}  \label{c-matrix}
\end{equation}%
where $C(t)$\ is the time-dependent \textit{collapse matrix}. We note that%
\begin{equation}
lim_{t\rightarrow \tau _{c}}C(t)=\sqrt{2}e^{i\phi }|r><r|\text{,}
\label{proyector}
\end{equation}%
being $|r><r|$ the standard Von Neumann projector. In our model, the
collapse matrix is some sort of time-dependent projector.

Considering energy conservation during transition $|+>\rightarrow |r>$, the
total energy radiated $E_{rad}$ by an electron is $\Delta E/2$, ($\Delta
E=E_{+}-E_{-}$). However, the electron may interchange energy $\Delta E_{xc}$
with the point contact during the measurement process. In such a case, we
note that both quantities ($\Delta E_{xc}$ and $\Delta E$) should be of the
same order of magnitude. If the interchange energy is higher than the level
splitting between both quantum wells, the resonant condition is not
obtained, and then, both symmetric $|\pm >$ pure states are destroyed.
Accordingly, the particle may interchange energy with an efficient
measurement device if $\Delta E_{xc}$ $\sim $ $\Delta E$. Taking into
account this, in this work we assume that $E_{rad}$ $\sim $ $\Delta E$. The
total energy radiated by an electron during measurement process $E_{rad}$ is 
\begin{equation}
E_{rad}=\int_{0}^{\tau _{c}}I(t)dt=\frac{e^{2}x_{r}^{2}}{2\pi \epsilon
c^{3}\tau _{c}^{3}}  \label{eradtotal}
\end{equation}%
where $t_{i}=0$ and $t_{f}=\tau _{c}$. Now we consider the tunneling process
between both quantum dots. We choose as our initial state $|l>$ and monitor
the time the particle takes to tunnel to the right-hand well when its state
is $|r>$. This tunneling process occurs under free evolution, and the time
taken is given by the inverse energy difference between the states $|+>$ and 
$|->$, $\tau _{t}=\pi \hbar /\Delta E$ \cite{cruz}. Then, and from Eq. (\ref%
{eradtotal}), the collapse time can be easily obtained 
\begin{equation}
\tau _{c}=\left( \frac{e^{2}x_{r}}{2\pi ^{2}\epsilon \hbar c^{3}}\tau
_{t}\right) ^{1/3}\text{.}  \label{tiempo}
\end{equation}%
\qquad \qquad

In Fig. 2, we have plotted the probability density in the right quantum well 
$|<r|\psi >|^{2}$ versus time during the measurement process. We have
considered a GaAs/Ga$_{1-x}$Al$_{x}$As double quantum dot system which
consists of two 150\AA -wide GaAs quantum wells separated by a barrier of
thickness $d=$15\AA\ and $d=$35\AA . The barrier height and electron
effective mass are taken to be 220meV and 0.067$m_{0}$, respectively. In
Fig. 3 and 4 is shown both collapse and tunneling times versus barrier
thickness. It is found that collapse times are three orders of magnitude
lower than tunneling times. We should point out that the classical electron
velocities associated with these collapse times are non-relativistic. In
Fig. 4, it is clearly shown that collapse time is increased as we increases
the barrier thickness. Such a result can be easily explained as follows. If
the barrier thickness is increased, the level splitting between both quantum
wells is decreased \cite{cruz} and both $\tau _{t}$ and $\tau _{c}$ are also
increased.

Finally, we think that the emission of electromagnetic radiation during
collapse process can be observed experimentally \cite{fu}. Each time a
measurement is realized in a double quantum dot system, an emission of an
electromagnetic radiation pulse takes place. The measured pulse width will
correspond to the collapse time value.

In summary, in this work we have considered an electron in a double quantum
dot system. We present a dynamical model for wave function collapse in the
strong coupling to the detector limit. A leats emitted radiation principle
give us the minimum possible collapse times without radiation reaction
divergences. Such collapse time values will correspond to minimum possible
measurement time values in a double quantum dot system.

\section{Figures}

\begin{itemize}
\item \textbf{Fig. 1} A schematic illustration of the coupled quantum dot
system. Conduction band potential and wave function amplitude of both $|r>$
and $|+>$ states.

\item \textbf{Fig. 2} Probability density in the right quantum well $%
|<r|\psi >|^{2}$ versus time during the measurement process. Thin line: 15%
\AA\ barrier thickness. Thick line: 35\AA\ barrier thickness.

\item \textbf{Fig. 3} Tunnelling time versus barrier thickness.

\item \textbf{Fig. 4 }Collapse time versus barrier thickness.
\end{itemize}

\end{document}